**Usability of back support, shoulder support and sit-stand passive occupational exoskeletons: A heuristic evaluation of the designs**

Alejandra Martinez, Laura Tovar, Carla Irigoyen Amparan, Karen Gonzalez, Prajina Edayath, Priyadarshini Pennathur, and Arunkumar Pennathur*

Physical, Information and Cognitive Human Factors Engineering Research Laboratory
Industrial, Manufacturing and Systems Engineering Department
University of Texas at El Paso, El Paso, Texas 79968-0521, USA

*Corresponding author; Email for corresponding author: apennathur@utep.edu

**Abstract**

Occupational exoskeletons promise to alleviate musculoskeletal injuries among industrial workers. Knowledge of the usability of the exoskeleton designs with respect to the user-device interaction points, and the problems in design features, functions and parts, evaluated and rated using design principles is still limited. Further, the usability of exoskeletons when assembling, donning, doffing and disassembling them, tasks that can be considered pre- and post-use tasks are also critical to evaluate, especially from a device design standpoint. We conducted a heuristic evaluation of the usability of three popular exoskeletons – a back support device, a shoulder support device, and a sit-stand exoskeleton when assembling, donning, doffing and disassembling them. Seven evaluators used Nielsen's and Shneiderman's usability heuristics to evaluate the devices. Results indicate that none of the three exoskeletons had any catastrophic usability problems, but all three had major usability problems including accommodating diverse users, the assembly, donning and doffing being a two person operation, poor documentation, a lack of sequence indicators during assembly of the devices, presence of safety hazards while donning and doffing the devices, and manual strength requirements. Further, the assembly task is the most difficult task resulting in the most violations of usability heuristics. The exoskeleton-human factors research community should include diverse users in their evaluations and conduct usability, accessibility, and safety evaluations of these devices to provide design feedback to device designers.

**Keywords:** exoskeleton, usability, heuristic evaluation, assembly, donning

**Relevance to Industry:** For workers in industry to widely adopt exoskeletons, exoskeleton manufacturers must design for a wider range of users, address critical safety concerns when assembling and donning these devices, and simplify designs to be a one-person operation.

# 1. Introduction

Occupational exoskeletons have emerged as a promising solution to alleviate the problem of work-related musculoskeletal disorders and maintain worker productivity and safety while performing industrial tasks (Elprama et al.,2022; Howard et al., 2020; Kermavnar et al., 2021; Kim et al., 2018; Kuber et al., 2022, 2023; McFarland & Fischer, 2019; Medrano et al., 2023; Nussbaum et al., 2018; Papp et al., 2020; Reid et al., 2017). But, for these devices to gain widespread acceptance among workers, and for workers to use these devices effectively to support their task performance, these devices must be comfortable and usable for prolonged wear during work shifts.

Studies investigating how exoskeletons assist with task performance, have also surveyed users whether the exoskeleton was comfortable when performing specific industrial tasks (Amandels et al., 2018; Antwi-Afari et al., 2021; Chae et al., 2021; Daratany & Taveira, 2020; Flor et al., 2021; Hensel & Keil, 2019; Kim et al., 2019; Smets, 2019; Winter et al., 2019; Ziaei et al., 2021), and whether the perceived exertion in the task is reduced due to exoskeleton use (Bock et al., 2021; Daratany & Taveira, 2020; De Vries et al., 2021; Flor et al., 2021; Kim et al., 2018; Liu et al., 2018; Siedl & Mara, 2021; Winter et al., 2019), in addition to assessing worker acceptance and intention to use the exoskeleton for their work (Flor et al., 2021; Gilotta et al., 2019; Hensel & Keil, 2019; Kim et al., 2018; Pacifico et al., 2022; Siedl & Mara, 2021; Spada et al., 2017; Turja et al., 2022). These user evaluation studies provide useful information on comfort, and user acceptance, with a focus on assessing whether the exoskeletons helped users perform the specific occupational task successfully with minimal discomfort and whether users exhibited intent to use and whether they would see themselves using these devices during their work.

These user evaluation studies also provide broader, overall perceptions of usability of the entire exoskeleton using tools such as System Usability Scale (SUS) questionnaire, and ad hoc usability questionnaires, but what would accelerate design improvements and make these devices usable are granular, specific details about the usability of the design features in these devices, the specific parts in the device and how these parts interact and impact the usability, and the user-device interaction points, and how the design of these interaction points hinder or promote usability. What is needed are design evaluations as a complement to what we know from user evaluation studies of exoskeletons. The level of detail about usability problems we will obtain from a device design evaluation will be different and will add to knowledge obtained from user evaluation studies. The type of information and evidence obtained about usability problems is substantially different in a design evaluation and can provide usability evaluation outcomes to improve the design of these devices.

Design evaluations focus on specific features such as the displays and the control points in the exoskeleton that the user interacts with – what we term the *user-device interaction points*. As examples, these user-device interaction points could include visual labels on the device instructing the user, a Velcro strap the user has to wear around them, a locking mechanism consisting of a slot in which the user must insert a metal part and lock using a turn knob, a sliding lever a user has to activate to determine size and fit, and so on. Evaluating the user-device interaction points using heuristic evaluations with usability design principles such as whether users are provided feedback on their actions when they interact with a control, whether users can recover from any mistakes they make when interacting with the device, whether the design of the device prevents user errors in the first place, and whether the design relies on the user to remember information or if the design only requires the user to recognize what they must do when they interact, etc., can generate valuable insights on design weaknesses at the user-device interaction points. Heuristic evaluations of usability can uncover unique and different usability problems compared to user evaluations during task performance. They provide granular details about problems in design features, functions and parts, and because human factors experts identify, evaluate, and rate the severity of the usability problems using design principles, designers can readily use the resulting prioritized set of usability problems to improve their designs.

Furthermore, most studies only focus on evaluating exoskeletons for and when performing industrial tasks, but there are pre- and post-use tasks that are also critical to evaluate, especially from a device design standpoint. Pre-use tasks include assembling the exoskeleton and donning it. Post-use tasks include doffing the exoskeleton, disassembling it for cleaning/sanitizing, and storing for next use. Unless the design of the exoskeleton supports the user-device interaction points for these pre-use and post-use tasks, the exoskeleton will become difficult to assemble, don, doff and disassemble and store for next use. If the device cannot be assembled, it cannot be donned, and used for the task. The design of the interaction points should also allow the worker to quickly doff the device (especially during the work shift if needed, during any emergencies, and at the end of the shift).

Given the gaps in knowledge about usability of exoskeletons from a design perspective during pre-use and post-use tasks, we conducted a heuristic evaluation of usability of three popular exoskeletons – a back support device, a shoulder support device, and a sit-stand exoskeleton when assembling, donning, doffing and disassembling them. In this paper, we report results from our study and discuss the implications for designing exoskeletons in the future.

## 2. Methods
### *2.1. Study Approach*

The study was divided into five phases. In the first phase, the evaluators met and reviewed the heuristics in detail, and discussed the evaluation process and the criteria and how the evaluation spreadsheets were to be completed. The exoskeletons were also set up in the laboratory for the evaluations. In the second phase, the project lead provided the evaluation spreadsheets to each evaluator, who then independently identified usability problems in each exoskeleton along with the heuristics each problem violated. The third phase consisted of combining the usability problems identified by the evaluators into categories of problems, and assigning labels to the categories of usability problems. In the fourth phase of the project, an individual spreadsheet for each exoskeleton was sent out, one a day, to each evaluator to rate the severity of the categorized and labeled usability problems they had identified. The completed spreadsheets were then sent to the lead every day by each evaluator, until all three exoskeletons had been rated by each evaluator independently. In the last phase of the project, the lead created a master spreadsheet with all the usability problems, and all the individual severity ratings and based on threshold criteria (discussed in the following sections), decided if a discussion was warranted. Discussions were then conducted among the evaluators, and the new ratings were recorded based on the discussion. The specific activities of the study process are shown in Figure 1. The study was completed in approximately four months.

***2.2. Exoskeletons Evaluated in the Study***

In this study, we evaluated three commercially available passive occupational exoskeletons: a shoulder support device, a lumbar flexion support device, and a sit-stand support exoskeleton. The shoulder support exoskeleton reduces fatigue and provides support to the worker when repetitive above shoulder work is performed. The back support device is intended to reduce and prevent back injuries. The sit-stand support exoskeleton allows a quick and seamless postural change between sitting, standing, and walking. Although the purpose of each device is different, and they all target different muscle groups, the overall goal of each exoskeleton is the same: to reduce and prevent injuries while supporting workers.

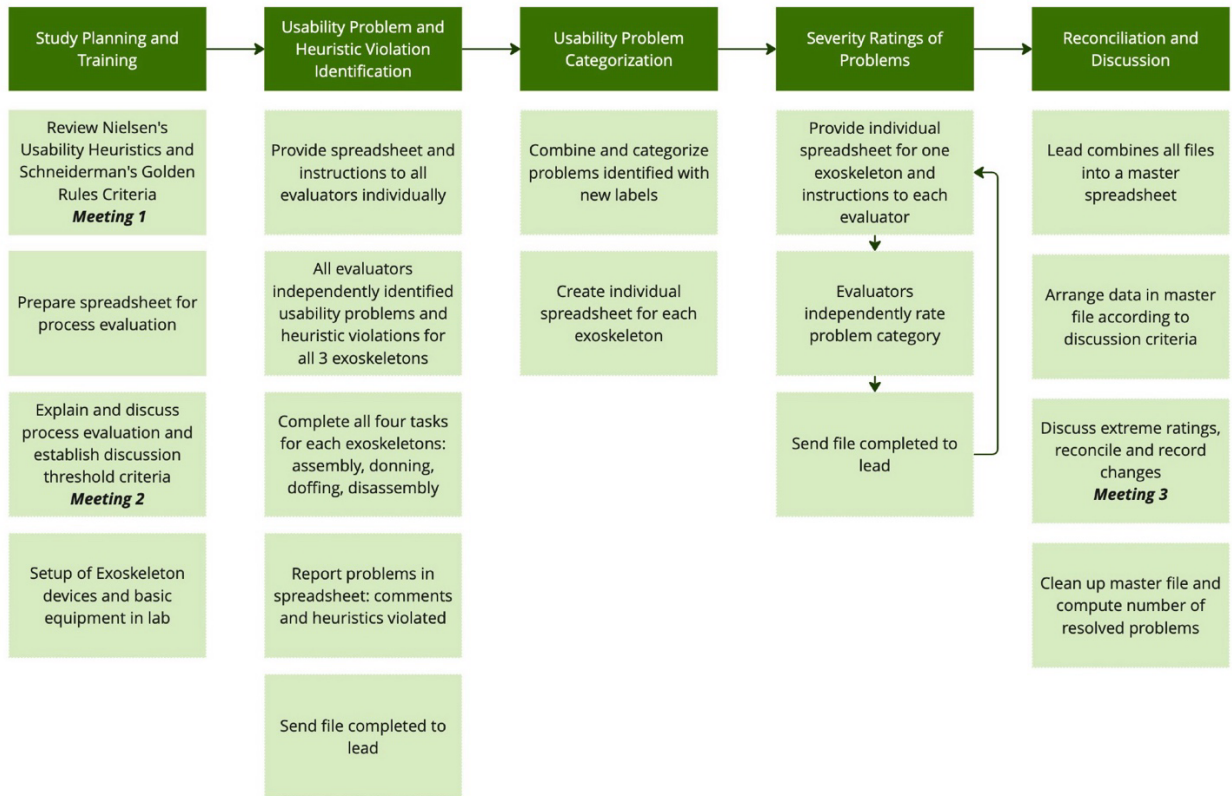

**Figure 1.** The major steps involved in the five study phases to complete the heuristic evaluations.

### 2.3. Usability Heuristics and Evaluation Criteria Used

In this study, we combined two sets of usability heuristics to evaluate the three passive occupational exoskeletons: Nielsen's usability heuristics (Nielsen, 1994, 2005) and Schneiderman's golden rules for interface design (Shneiderman et al., 2016). Heuristic evaluation is a usability inspection technique and is considered to be more of an informal usability evaluation method given it relies on use of heuristics. But compared to empirical usability techniques such as testing with real users, and more formal techniques such as a complete task analysis with users, heuristic evaluation is easy to use, easy to learn, and can help identify major usability problems in a product in a quick and cost-effective way, with very little time commitment required of the evaluators (Klarich et al., 2022; Mack & Nielsen, 1995;

Martinez et al., 2024; Nielsen & Molich, 1990; Zhang et al., 2003). Furthermore, our focus was on evaluating the usability problems related to the specific design weaknesses of the exoskeletons, using our human factors engineering knowledge and expertise, for which heuristic evaluation was a suitable option.

In our study, 14 heuristics, ten from Nielsen and four from Shneiderman's usability heuristics, were used. Table 1 presents the 14 evaluation heuristics, the criteria for each heuristic and the evaluation guidelines for each heuristic in the form of questions to facilitate evaluation. To evaluate each heuristic, all evaluators were asked to use the same severity rating scale, and before rating the severity of the problem, to consider if the problem would be a persistent problem every time a user used the product, or if it would only be a one-time or a first time usability problem that would not be a problem for the user with repeated use. Table 1 represents the severity rating scale used.

**Table 1.** Severity rating scale used for the heuristic evaluation of the three exoskeletons.

| Severity Rating | Rating Description |
|---|---|
| 0 | I don't agree that this is a usability problem at all. |
| 1 | Cosmetic problem only; need not be fixed unless extra time is available. |
| 2 | Minor usability problem; fixing this should be given low priority. |
| 3 | Major usability problem; important to fix, so should be given high priority. |
| 4 | Usability catastrophe; imperative to fix this problem. |

**Table 2.** Design principles for Nielsen's Usability Heuristics and Shneiderman's golden rules for interface design, its criteria and evaluation questions for rating.

| Design principle | Criteria | Ask each question below and then evaluate with your rating |
|---|---|---|
| Visibility of system status | Designs should keep users informed about what is going on, through appropriate, timely feedback. | 1. Does the design provide the user information on the current state of the system?<br>2. Does the design provide the user information on what actions can be taken by the user at the current state?<br>3. Does the design indicate to the users what the next step will be?<br>4. Does the design indicate what changes occur after the user takes an action?<br>5. Is feedback presented quickly to the user after their actions? |
| Match between system and real world | The design should speak the users' language and use words, phrases, and concepts familiar to the user and the image of the system should match the model the users have about the system | 1. Will users be familiar with the terminology used in the design?<br>2. Do the design's controls follow real-world conventions?<br>3. Does the user model match system image?<br>4. Do actions provided by the design match actions performed by users?<br>5. Do objects in the design match objects in the task? |
| User control and freedom | Users often perform actions by mistake. They need a clearly marked "emergency exit" to leave | 1. Does the design allow users to go back one step in the process?<br>2. Are exit points easily discoverable?<br>3. Can users easily cancel an action they took?<br>4. Is Undo and Redo supported?<br>5. Are any actions users have to take surprising?<br>6. Do any actions lead to unexpected outcomes?<br>7. Are the actions tedious for the user? |
| Consistency and standards | Users should not have to wonder whether different words, situations or actions mean the same thing. Follow platform conventions. | 1. Does the design follow industry conventions in color, layout and position, font, capitalization, terminology and language, and any standards?<br>2. Are the sequences of user actions consistent?<br>3. Are visual treatments used consistently throughout the design? |
| Error prevention | Good error messages are important, but the best designs carefully prevent problems from occurring in the first place | 1. Does the design have interfaces that make errors impossible?<br>2. Does the design prevent errors of evaluation and execution?<br>3. Does the design prevent slips and mistakes by using helpful constraints?<br>4. Does the design warn users before they perform risky actions? |
| Recognition rather than Recall | Minimize the user's memory load by making elements, actions and options visible. Avoid making users remember information. | 1. Does the design keep important information visible so that users do not have to memorize it?<br>2. Does the design offer contextual help? |
| Flexibility and efficiency of use | Shortcuts, hidden from novice users, may speed up the interaction for the expert user. Users always learn and users are always different. Give users the flexibility of creating customization and shortcuts to accelerate their performance. | 1. Does the design provide shortcuts for expert users?<br>2. Is the content and functionality personalized or customized for individual users? |
| Aesthetic and minimalist design | Interfaces should not contain information which is irrelevant. Every extra unit of information in an interface competes with relevant units of information. | 1. Is the visual design and content focused on the essentials?<br>2. Have all distracting, unnecessary elements been removed? |
| Recognize, diagnose and recover from errors | Error messages should be expressed in plain language. | 1. Does the design use traditional error message visuals like bold, red text?<br>2. Does the design offer a solution that solves the error immediately?<br>3. Do the error messages allow the user to understand the nature of errors, learn from errors, and recover from errors? |

| | | | |
|---|---|---|---|
| | | 4. | Are the error messages precise and not vague or general? |
| | | 5. | Are the error messages constructive and polite? |
| | | 6. | Does the design allow users to recover from errors at various levels - a single action, a subtask, or a complete task? |
| | | 7. | Does recovery from errors require multiple steps? |
| | | 8. | Does recovery from errors encourage exploratory learning? |
| Help and documentation | Its best if the design doesn't need any additional explanation. However, it may be necessary to provide documentation to help users complete their tasks. | 1. | Is help documentation available and easy to search and use? |
| | | 2. | Is help context sensitive and provided at the moment the user requires it? |
| | | 3. | Is help task-oriented or alphabetically ordered, or semantically organized, or searched? |
| Seek universal usability | Recognize the needs of diverse users and design for plasticity. | 1. | Does the design accommodate novices to experts, different age ranges, disabilities, international variations, and technological diversity? |
| Offer informative feedback | For every user action, there should be an interface feedback. For frequent and minor actions, the response can be modest, but for infrequent and major actions, the response should be substantial. | 1. | Does the design show changes and feedback from user actions visually? |
| Design to yield closure | Action sequences should be organized into groups with a beginning, middle and end. | 1. | Does the design suggest a clear beginning, middle and end in using the product? |
| | | 2. | Does the design provide clear feedback to the user that user goals are achieved in these stages? |
| Use users' language | The language should be always presented in a form understandable by intended users | 1. | Does the design use standard meanings of words? |
| | | 2. | Does the design use specialized language for specialized groups? |

### *2.4. Tasks evaluated in the study*

Our study was designed to heuristically evaluate usability of the exoskeletons while assembling, donning, doffing, and disassembling the devices. To evaluate usability of a device when assembling it, the evaluator had to use all the necessary parts and put together the complete exoskeleton in all possible configurations – for instance, the sit-stand device could be worn in multiple ways – either with a waist belt or with a shoulder belt or with both, hence the evaluator had to evaluate usability for all 3 configurations of the device. To evaluate usability of the device during the donning task, the evaluator had to don the exoskeleton in its entirety for it to be counted as a successful attempt, to replicate the real world scenario of donning the exoskeleton as though the device was going to be used for work. The doffing task involved evaluating the usability of successfully removing the complete exoskeleton from the evaluator's body. Lastly, to evaluate usability when disassembling the device, the evaluator had to disassemble the exoskeleton successfully until it was in its original configuration ready for the next assembly task sequence. The evaluators were instructed to take notes if they were unsuccessful in any task phase. All evaluators were free to use all available tools such as the instruction manuals that came with the exoskeletons, more detailed online manuals accessed via a computer, and hand tools such as Allen wrenches that came with the exoskeletons.

## 2.5. Heuristic evaluation process and procedure

After reviewing all the usability principles, heuristics and the evaluation criteria, the evaluation process and establishing a threshold discussion criterion (Figure 1), the team lead sent out the evaluation spreadsheet to all other evaluators for them to record all their work in the same format and in one single place. The evaluators were instructed to identify and specify the usability problem, the heuristics that were violated for the problem, and add their recommendations, if any. The evaluators had to complete these steps when assembling, donning, doffing and disassembling each exoskeleton. A snippet of the spreadsheet sent out can be seen in Figure 2.

**Figure 2:** Screenshot of the first spreadsheet template used to complete heuristic evaluations.

After all the evaluators completed the evaluation process they sent their completed spreadsheet to the team lead. The spreadsheets were then compiled into one master file which contained the original columns (Violated?, Problems, Heuristics violated, Recommendations) of each evaluator and categorized by the assembly, donning, doffing and disassembly tasks.

The master file contained all the usability problems documented by each evaluator. These problem statements were then analyzed to create problem categories based on similar descriptions of problem statements from each evaluator. Each problem category then had a 'usability problem category label' assigned to it (see figure 3 for an example). This process of problem categorization and labeling was done for all the exoskeletons evaluated and three new files, one corresponding to each exoskeleton, were created. The new files contained the name of the task, the usability problem category, a problem number, the usability problem statements from the evaluators and the heuristics that the evaluators reported were violated when experiencing a usability problem. The last column was specific to each evaluator, and in this column, they were instructed to record their severity rating of the usability problem represented by the problem category, in the same row as the problem category label. Evaluators were instructed to complete these severity ratings individually, and while completing the severity ratings, to think if the problem would be a recurrent, persistent problem or just a one-time problem.

The spreadsheet for one exoskeleton was sent out at the beginning of the day by the team lead to each evaluator. The evaluators were instructed to return the spreadsheet file at the end of the day. After spreadsheets for all 3 exoskeletons were returned to the team lead by all evaluators, the lead compiled the final master file with all the ratings (see figure 4).

**Figure 3:** Screenshot of second evaluation template used for assigning a severity rating of the usability problem.

**Figure 4:** Screenshot of final master file with the severity ratings of all evaluators; this image pertains to one exoskeleton only, the master file contains different tabs for each exoskeleton, all with the same format.

## 2.6. Reconciliation and Discussion Process

We developed discussion threshold criteria (see Figure 1) to determine items for discussion after evaluation and ratings were complete. This threshold criteria involved a set of rules established by the team to discuss the problems with extreme ratings to resolve conflicts. For the Nielsen and Shneiderman design principles a severity rating of 0 indicated that there was no problem at all with the design and a severity rating of 4 indicated a catastrophe. Figure 5 illustrates the criteria utilized to determine which problems would be discussed. This threshold was applied to the severity ratings data for all exoskeletons to facilitate discussion.

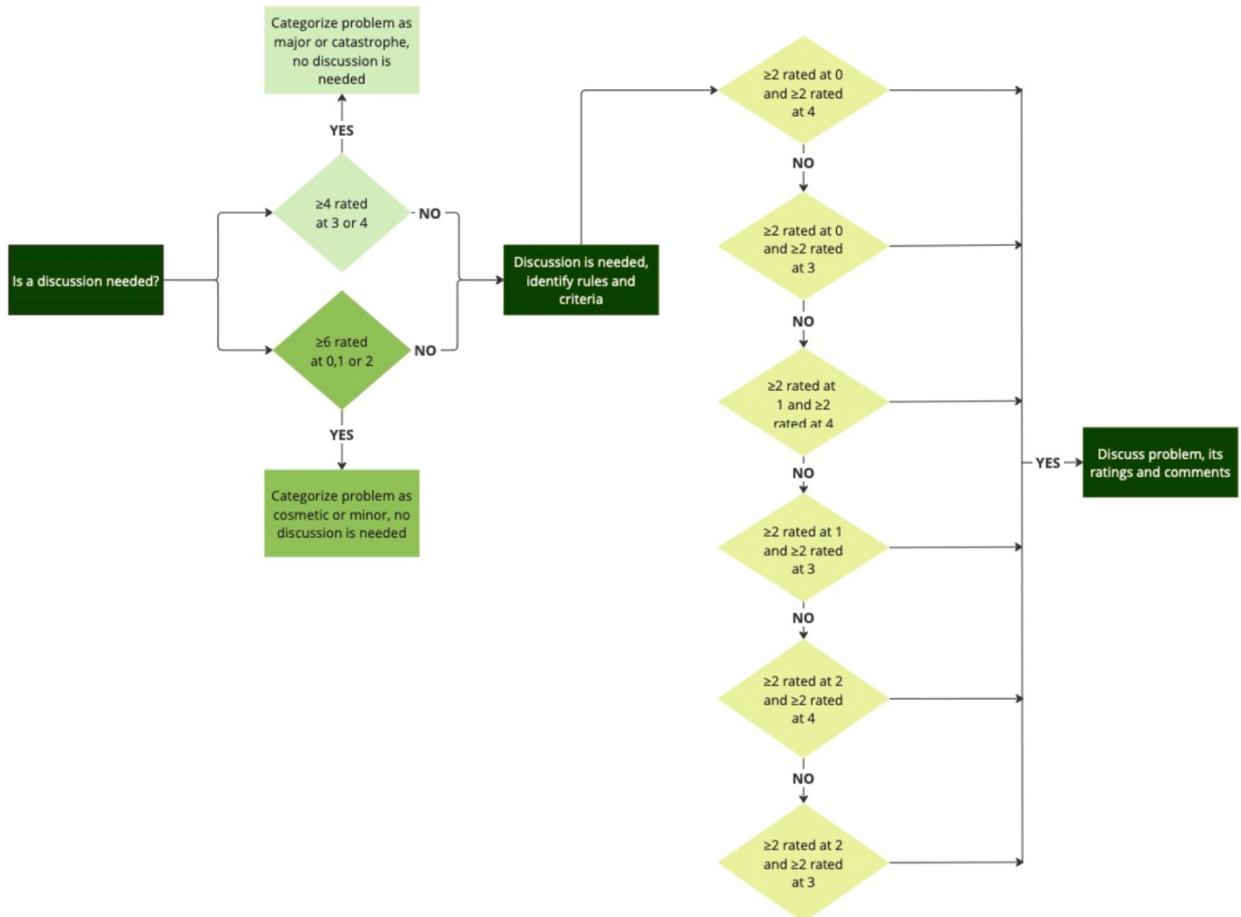

**Figure 5:** Decision process used to determine if a problem needed discussion using established discussion threshold criteria.

For usability problem categories that required discussion, the evaluators with extreme ratings explained their reasoning for their severity rating first. Other evaluators then joined the discussion. After hearing from all the evaluators, each evaluator then decided to either keep their rating or provide a new rating. The project lead recorded all the changes to the severity ratings made during the meeting, and once all the problems were discussed, the number of split ratings that converged were computed after one round of discussion. Table 3 lists the total number of criteria discussed per exoskeleton and the number of criteria that converged after discussion. For five criteria, evaluators moved their ratings up or down slightly based on the discussion. For the lumbar flexion support exoskeleton, the average rating for the one problem discussed changed from 2.2 before discussion to 2.4 after discussion. For the shoulder support device, the average ratings changed from 1.57 to 2.2 for one problem and remained at 2.4 for the second problem discussed. For the sit-stand support device, the average ratings decreased from 2.14 to 2.0, and increased marginally from 1.57 to 2.0 for the two problems discussed. These new average ratings continued to reflect only a minor usability problem in these devices.

**Table 3:** Total number of criteria discussed for all exoskeletons and the number of criteria that converged after discussion and application of threshold criteria.

| Type of Exoskeleton | Number of criteria discussed | Number of reconciled and converged criteria after discussions |
|---|---|---|
| Lumbar flexion support exoskeleton | 9 | 8 |
| Shoulder support exoskeleton | 17 | 15 |
| Sit-stand support exoskeleton | 20 | 18 |

### *2.7. Statistical analysis*

The following descriptive statistical analyses were conducted for this study:

a. The frequency with which a usability heuristic was violated across the four tasks (assembly, donning, doffing, and disassembly).

b. The binning of usability problems by their mean rating across all tasks: The mean severity rating was calculated for each problem. Then, based on the average severity rating of the problem, the problem was classified as being cosmetic, minor, major and catastrophic, and a count was generated for each bin (cosmetic problem, minor problem, major problem, catastrophic problems) for each exoskeleton.

c. The raw severity rating score of each problem organized by the four tasks: The severity ratings for all evaluators were represented as a boxplot for better visualization of how individual evaluators rated each usability problem across the assembly, donning, doffing, and disassembly tasks for each exoskeleton.

d. The number of times a general, major usability problem was repeated across assembly, donning, doffing, and disassembly tasks across the three exoskeletons: The major (2.5 to 3.5 average severity ratings) usability problems were first identified and listed in a spreadsheet for each exoskeleton for each task phase. Then, the major usability problem labels were examined and compared across exoskeletons and task phases to identify problems that were present across exoskeletons, and across task phases. These were classified as repeated major usability problems. Problems that only occurred in any one exoskeleton were classified as unique problems. The repeated and unique major usability problems were then listed in a table and their frequencies counted.

e. The count of the times a heuristic was violated in each of the repeated, general, major usability problems across the three exoskeletons in the four tasks completed.

## 3. Results

Our research questions aimed to identify usability problems in four tasks namely assembling, donning, doffing, and disassembling each exoskeleton and the usability heuristics that were violated the most in the 3 exoskeletons we evaluated. Additionally, we sought to identify catastrophic usability problems, and major usability problems, if any, in each of these exoskeletons, because catastrophic and major usability problems with average severity ratings between 2.5 and 4 require immediate priority and attention from the exoskeleton designers and manufacturers for solving the problems and improving the designs. Because we evaluated three different types of exoskeletons in this study (a shoulder support, a back support and a sit-stand device), we were also interested in identifying if the catastrophic and major usability problems, if any, were unique and specific only to a particular exoskeleton, or if they were general usability problems that were repeatedly present in more than one exoskeleton or in more than one task phase – this would indicate if the design weaknesses underlying the usability problems were pervasive.

### *3.1. Overall trends in usability problems and heuristics violated*

Figure 6 displays the count of usability problems categorized by the average severity ratings of the problem. Following the schema for average severity ratings classification from Zhang et al. (2003) and Klarich et al. (2022) an average severity rating of 0 to 1.5 indicated a cosmetic usability problem, 1.5 to 2.5 indicated a minor problem, 2.5 to 3.5 indicated a major problem, and 3.5 to 4.0 indicated a catastrophic usability problem. A key finding from our analyses was that there were no catastrophic usability problems found for any of the 3 exoskeletons evaluated. However, major usability problems for the shoulder support exoskeleton accounted for nearly 44% of the total problems found in that exoskeleton – this was higher than minor (38%) or cosmetic problems (18%) in this device. The back support exoskeleton and the sit-stand exoskeleton had 37% and 35% major usability problems respectively, which were fewer than the number of minor problems found in these exoskeletons.

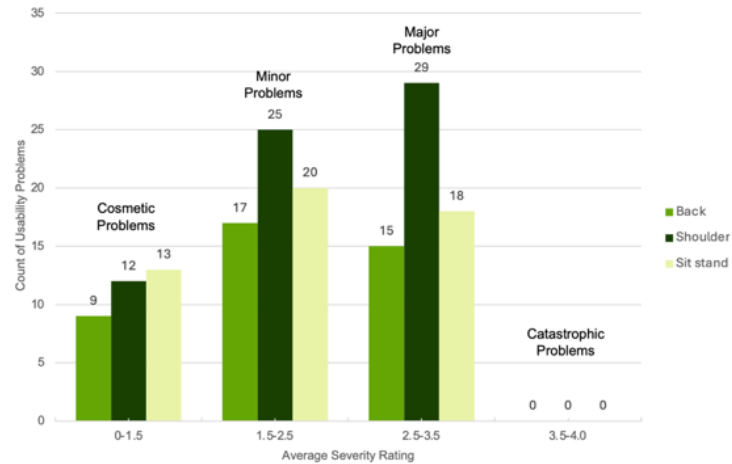

**Figure 6.** Number of usability problems categorized by average severity ratings for the 3 exoskeletons.

Figures 7, 8 and 9 show severity ratings for usability problems for the three exoskeletons. Figure 7 illustrates the severity ratings of all usability problems organized by the four task phases for the shoulder support exoskeleton. As can be seen from figure 7, the assembly task (panel a), had 26 problems, which was more than the donning (panel b), doffing (panel c), and disassembly tasks (panel d). Furthermore, the assembly task had the most cosmetic problems (a severity rating of 1 or less), particularly, in lacking shortcuts, and in providing visible information to users. Evaluators agreed to a great degree that a lack of visible information in the device was a cosmetic problem, with only one outlier severity rating of 3. Donning and disassembly tasks had no cosmetic problems.

Furthermore, figure 7 shows that the assembly task also had the highest number of major problems (at a severity rating of 3 or above), with 6 in total. The major problem of assembly being a two person operation had 100% agreement from the evaluators; further, during assembly, the major problems of accommodating different people, documentation, and force requirements for activating the arm cup had a high level of agreement in severity ratings from the evaluators. Disassembly was the only task without any major problems.

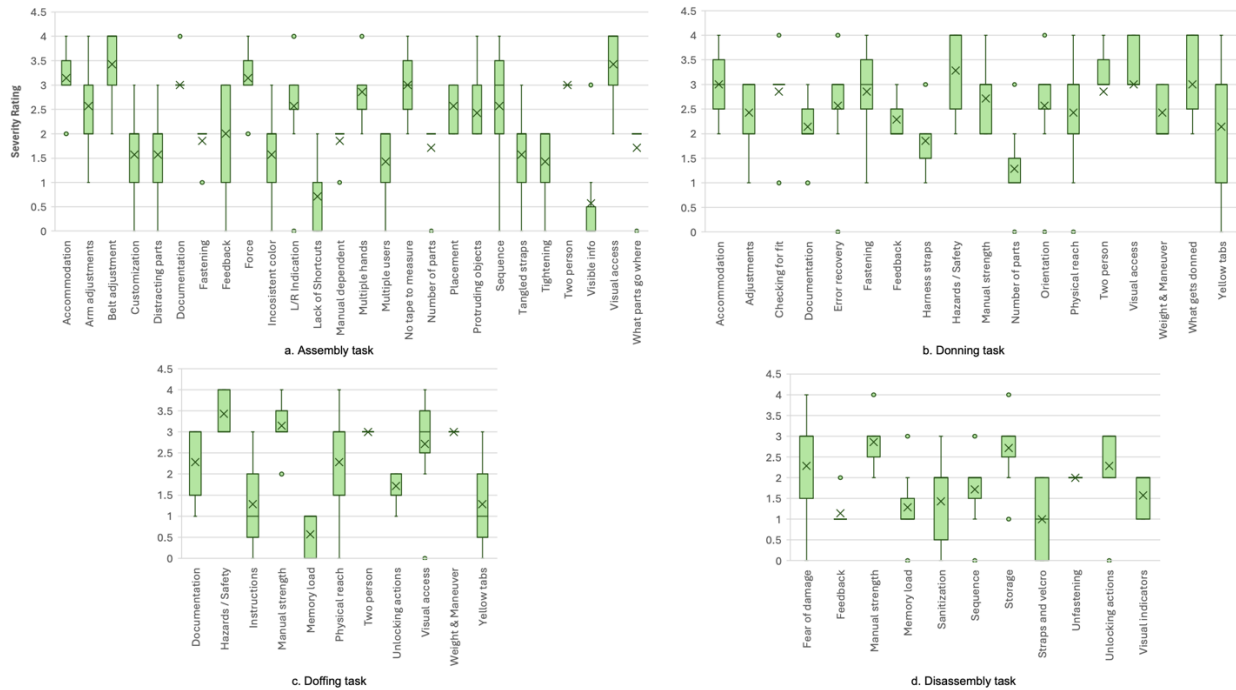

**Figure 7(a), 7(b), 7(c) and 7(d).** Boxplots for usability problems in the shoulder support exoskeleton organized by the four task phases.

Figure 8 presents the breakdown of the severity ratings of all usability problems by task phases for the back support exoskeleton. Evaluators found significantly fewer usability problems for the doffing task (panel c), whereas assembly (panel a), donning (panel b), and disassembly tasks (panel d) had about the same number of usability problems. It can also be seen from figure 8 that there were no cosmetic problems (at a rating of 1 or below) in the assembly, doffing and disassembly tasks. Evaluators further agreed that a lack of orientation and mapping for donning the device was the only cosmetic problem in this exoskeleton.

The evaluators found five major problems (at a severity rating of 3 or above) with a high level of agreement in the assembly task: a lack of stopping constraints, leg pad locking, a large number of parts, lack of sequence indicators and visual overload. No major problems were found during doffing.

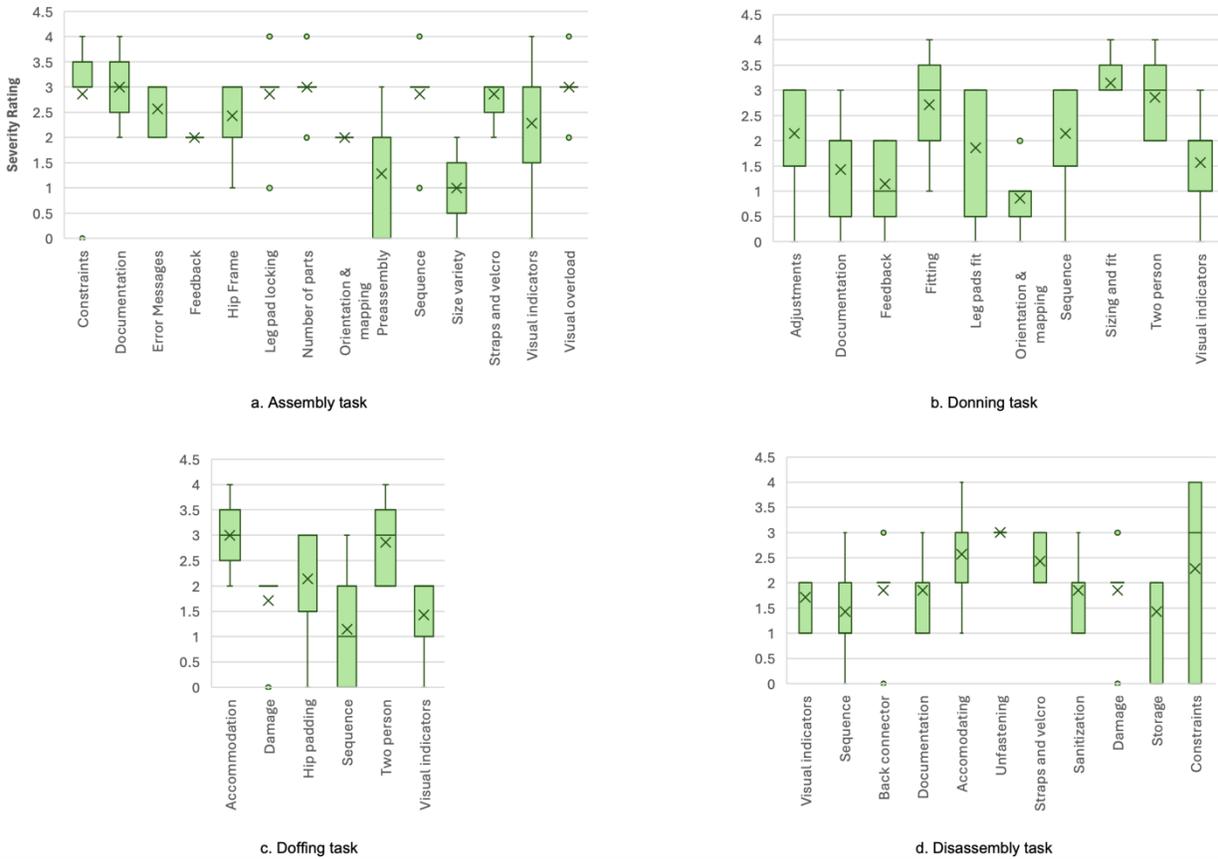

**Figure 8(a), 8(b), 8(c) and 8(d).** Boxplots for usability problems in the back support exoskeleton organized by the four task phases.

Figure 9 presents the breakdown of the severity ratings for all usability problems categorized by task phases for the sit-stand exoskeleton. We can see from figure 9 that the disassembly task (panel d) had significantly fewer usability problems than the assembly (panel a), donning (panel b) and doffing tasks (panel c). Evaluators agreed that the doffing task had the highest number of cosmetic problems – problems included evaluators being unable to mentally map and understand how the device and its component parts would have to be doffed.

Figure 9 also indicates that for the sit-stand exoskeleton, the donning task had the highest number of major usability problems. These problems included accommodating different people, size and weight matching people's ability to put the device on, and donning being a two-person operation. Additionally, requiring two people to operate the device was a problem for the assembly, donning, and doffing tasks, but received a high level of agreement in severity ratings from evaluators only for the donning and doffing tasks. The evaluators did not document any major usability problems when disassembling the sit-stand exoskeleton.

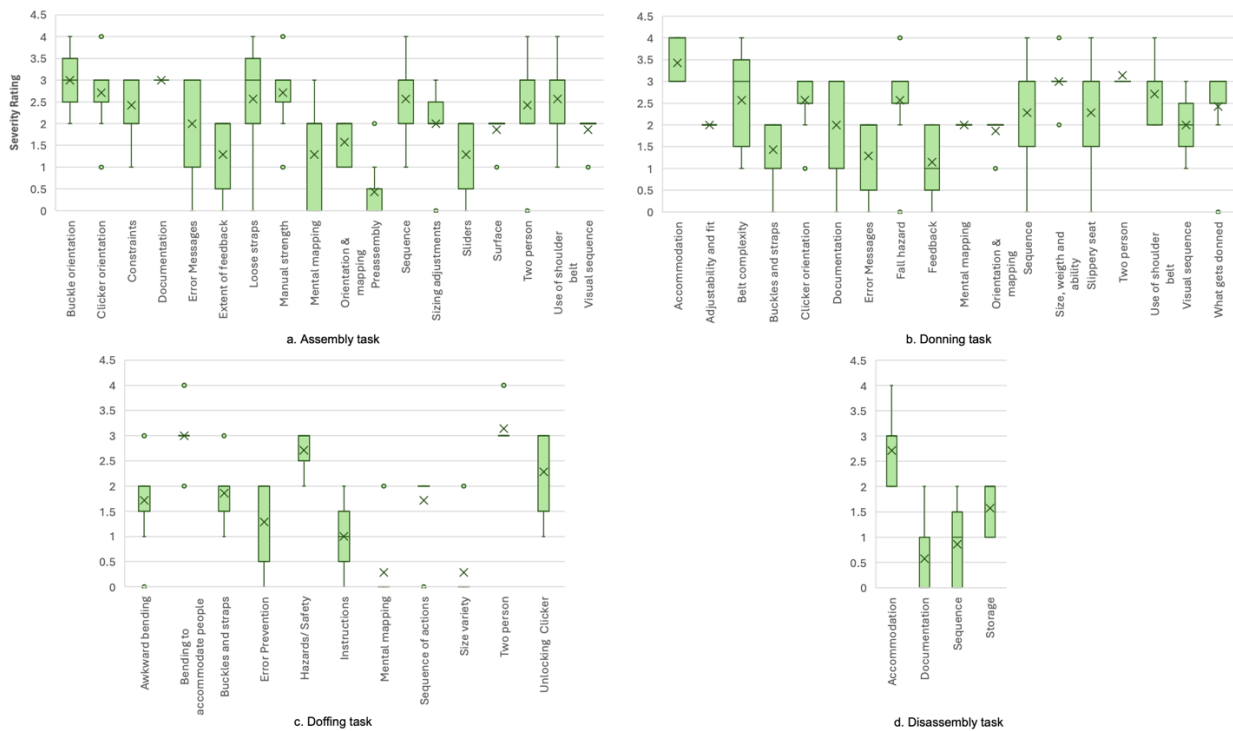

**Figure 9(a), (b), (c) and (d).** Boxplots for usability problems in the sit-stand support exoskeleton organized by the four task phases.

### *3.2. Usability problems and count of heuristics violated*

Figure 10 presents the number of times a usability heuristic was violated summed over all usability problems (top row) and summed over only the major problems (bottom row) for each exoskeleton. The visibility of system status, the match between system and the real world, user control and freedom, error prevention, and universal usability were among the top five most violated heuristics over all usability problems. This trend in violation was repeated in all the three exoskeletons. Also, the visibility of system status was violated the most for the back support exoskeleton and the sit-stand support exoskeleton, whereas user control and freedom was the most violated for the shoulder support exoskeleton. These findings change slightly when examining only the major usability problems. For the sit-stand support exoskeleton, the heuristic that assesses whether the design provided closure for the task had higher number of violations than the heuristic evaluating match between the system and the real world. However, the visibility of system status remained the most violated usability heuristic for this exoskeleton. For the shoulder support exoskeleton, user control and freedom also remained as the most violated usability heuristic. As for the back support exoskeleton, user control and freedom was the heuristic with most violations.

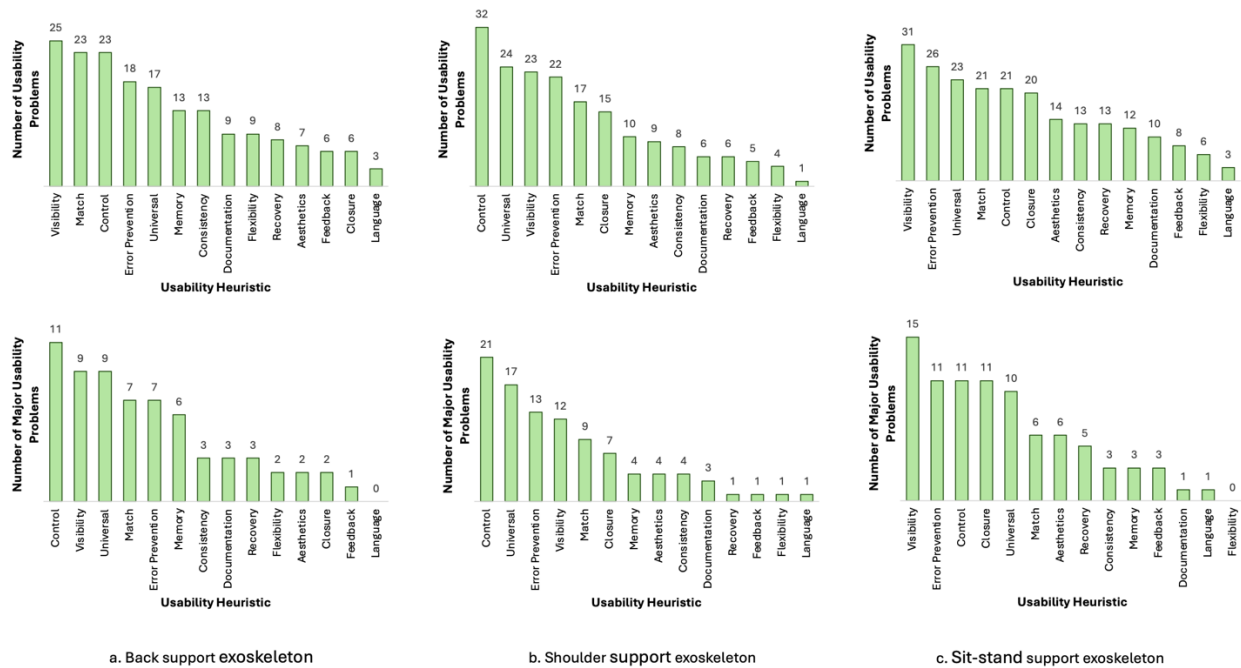

**Figure 10.** Count of heuristic violations measured by the number of all usability problems (cosmetic, minor and major), and only major problems for the 3 exoskeletons.

### *3.3. Major usability problems and heuristics violated for the 3 exoskeletons*

3.3.1. Overall trends

The count of heuristics violated for major usability problems (with severity ratings of 2.5 to 3.5) by each task phase for each exoskeleton (figure 11) indicates that across all 3 exoskeletons, most heuristics were violated when performing the assembly task. About 54% of the heuristics violated for major usability problems for the back support exoskeleton, 40% for the shoulder support exoskeleton, and 41% for the sit-stand exoskeleton pertained to the assembly task.

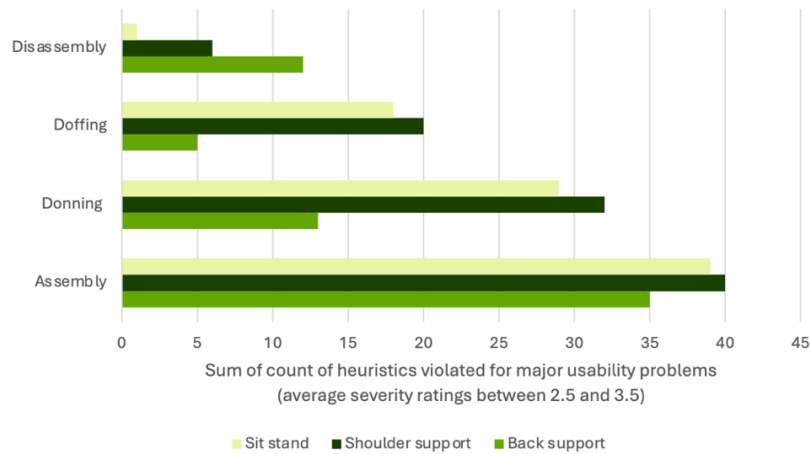

**Figure 11.** Count of heuristics violated for major usability problems by task phases for all exoskeletons.

3.3.2. Major usability problems in shoulder support exoskeleton

The shoulder support exoskeleton had 28 major usability problems, of which 15 were unique to this exoskeleton, and 13 were general problems that were found in at least one other exoskeleton (table 4). Figure 12a, b and c depict problems unique to the shoulder support exoskeleton, including a major problem with adjusting the belt during assembly (figure 12a) with a mean severity rating of 3.4, and safety hazards which had a mean severity rating of 3.4 for both the donning and doffing tasks. Limited visual access due to the color, size, shape or position of various components, inhibited the evaluators from seeing important components of the exoskeleton when assembling (mean rating of 3.4), donning (mean rating of 3.0) and doffing (mean rating of 2.7). The force required for activating the arm cup (figure 12b) was also rated as a unique major problem by the evaluators with a mean severity rating of 3.1.

Figure 12c portrays the major usability problem of weight and device maneuverability for doffing – the evaluators agreed that it was too difficult to maneuver this heavy device. This problem was present during the donning task as well (see figure 7b) which the evaluators rated as a minor problem, with a mean severity rating of 2.4. Manual strength requirement was a major usability problem for donning (depicted in the figure 12d with a mean severity rating of 2.9), doffing (a mean severity rating of 3.1), and disassembly (a mean severity rating of 2.9) for the shoulder support exoskeleton.

**Table 4.** Major usability problems organized by task phase (with average severity ratings between 2.5 and 3.5) in the shoulder support exoskeleton.

| Task phase | Major usability problem | Whether general or unique problem | Average severity rating |
|---|---|---|---|
| Assembly | Belt adjustment | Unique | 3.4 |
| | Limited visual access | Unique | 3.4 |
| | Force for arm cup activation | Unique | 3.1 |
| | Accommodating different people | General | 3.1 |
| | Two person operation | General | 3.0 |
| | Lack of measurement tools | Unique | 3.0 |
| | Documentation | General | 3.0 |
| | Multiple hands needed | Unique | 2.9 |
| | Placement of flex frames | Unique | 2.6 |
| | Arm length adjustments | Unique | 2.6 |
| | Sequence indicators | General | 2.6 |
| | Left/right indications | Unique | 2.6 |
| Donning | Hazards/safety considerations | General | 3.4 |
| | Accommodating different people | General | 3.0 |
| | Limited visual access | Unique | 3.0 |

|  | Fastening actions | Unique | 2.9 |
|  | What gets donned first? | General | 2.9 |
|  | Checking for fit | Unique | 2.9 |
|  | Manual strength requirements | General | 2.9 |
|  | Orientation of parts | Unique | 2.6 |
|  | Error recovery | Unique | 2.6 |
| Doffing | Hazards/safety considerations | General | 3.4 |
|  | Manual strength requirements | General | 3.1 |
|  | Two person operation | General | 3.0 |
|  | Weight and device maneuverability | Unique | 3.0 |
|  | Limited visual access | Unique | 2.7 |
| Disassembly | Manual strength requirements | General | 2.9 |
|  | Storage requirements | General | 2.7 |

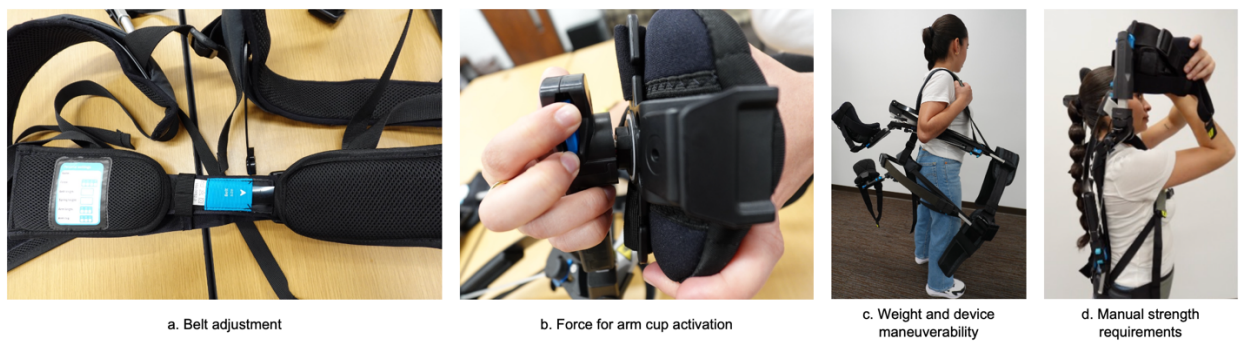

**Figure 12a, b, c, and d**. Major usability problems for the shoulder support exoskeleton organized from highest mean severity rating to lowest mean severity rating (left to right).

3.3.3. Major usability problems in back support exoskeleton

Table 5 depicts the major usability problems for the back support exoskeleton organized by task phases. There was a total of 15 major usability problems, of which 11 were general problems, with only 4 problems being unique to the design of this exoskeleton. The major usability problem of sizing and fit during the donning task received a severity rating of 3.1, the highest for this exoskeleton. Checking for fit required (at least for the first time) donning the device. Furthermore, assistance from another person was required for proper fitting (figure 13d) because the "check for fit" after donning needs to be performed at the user's back with no visual or physical access.

**Table 5.** Major usability problems organized by task phase (with average severity ratings between 2.5 and 3.5) in the back support exoskeleton.

| Task phase | Major usability problem | Whether general or unique problem | Average severity rating |
|---|---|---|---|
| Assembly | Number of parts | General | 3.0 |
| | Documentation | General | 3.0 |
| | Visual overload | Unique | 3.0 |
| | Straps and Velcro | General | 2.9 |
| | Leg pad locking | Unique | 2.9 |
| | Sequence indicators | General | 2.9 |
| | Stop constraints | General | 2.9 |
| | Lack of error messages | General | 2.6 |
| Donning | Sizing and fit | Unique | 3.1 |
| | Two person operation | General | 2.9 |
| | Fitting assistance | Unique | 2.7 |
| Doffing | Accommodating different people | General | 3.0 |
| | Two person operation | General | 2.9 |
| Disassembly | Tedious unfastening methods | General | 3.0 |
| | Accommodating different people | General | 2.6 |

Figure 13a showcases a major problem with the large number of parts in the exoskeleton hindering assembly of the back support exoskeleton. For disassembly, tedious unfastening methods (figure 13b) was also a top-ranked problem, with the same mean severity rating as the number of parts. Lastly, figure 13c depicts the leg pad locking problem due to the difficulty in attaching the kneecaps to the hip frame.

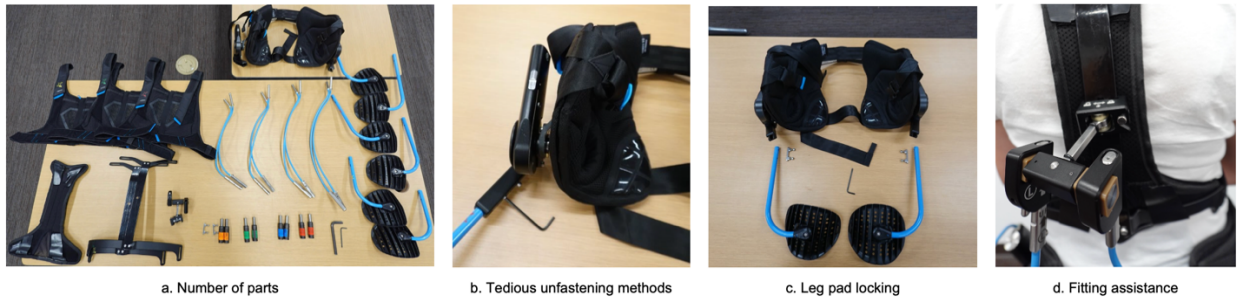

**Figure 13a, b, c and d.** Major usability problems for the shoulder support exoskeleton organized from highest mean severity rating to lowest mean severity rating (left to right).

3.3.4. Major usability problems in the sit-stand support exoskeleton

For the sit-stand support exoskeleton, evaluators identified a total of 18 major usability problems, of which 10 were unique to this exoskeleton, and 8 were general problems found in at least one more device (table 6). Accommodating different people was a major usability problem with a mean severity rating of 3.4, the highest for this exoskeleton. Evaluators also noted this problem during the disassembly task, with a 2.7 mean severity rating. In the donning task phase, the size, weight, and ability requirements (figure 14b) was a major problem.

**Table 6.** Major usability problems organized by task phase (with average severity ratings between 2.5 and 3.5) in the sit-stand exoskeleton.

| Task phase | Major usability problem | Whether general or unique problem | Average severity rating |
| --- | --- | --- | --- |
| Assembly | Buckle orientation | Unique | 3.0 |
| | Documentation | General | 3.0 |
| | Manual strength requirements | General | 2.7 |
| | Shoe clicker orientation | Unique | 2.7 |
| | Use of shoulder belt | Unique | 2.6 |
| | Sequence indicators | General | 2.6 |
| | Loose straps | Unique | 2.6 |
| Donning | Accommodating different people | General | 3.4 |
| | Two person operation | General | 3.1 |
| | Size, weight and ability requirements | Unique | 3.0 |
| | Use of shoulder belt | Unique | 2.7 |
| | Belt complexity | Unique | 2.6 |

|  | Shoe clicker orientation | Unique | 2.6 |
|---|---|---|---|
|  | Fall hazard | Unique | 2.6 |
| Doffing | Two person operation | General | 3.1 |
|  | Awkward bending postures | Unique | 3.0 |
|  | Hazards/safety considerations | General | 2.7 |
| Disassembly | Accommodating different people | General | 2.7 |

Figure 14a, c, and d all correspond to unique problems in the assembly task phase in the sit-stand exoskeleton, with buckle orientation and, size, weight and ability requirements with a mean severity rating of 3.0, and shoe clicker orientation and loose straps with a mean severity rating of 2.6.

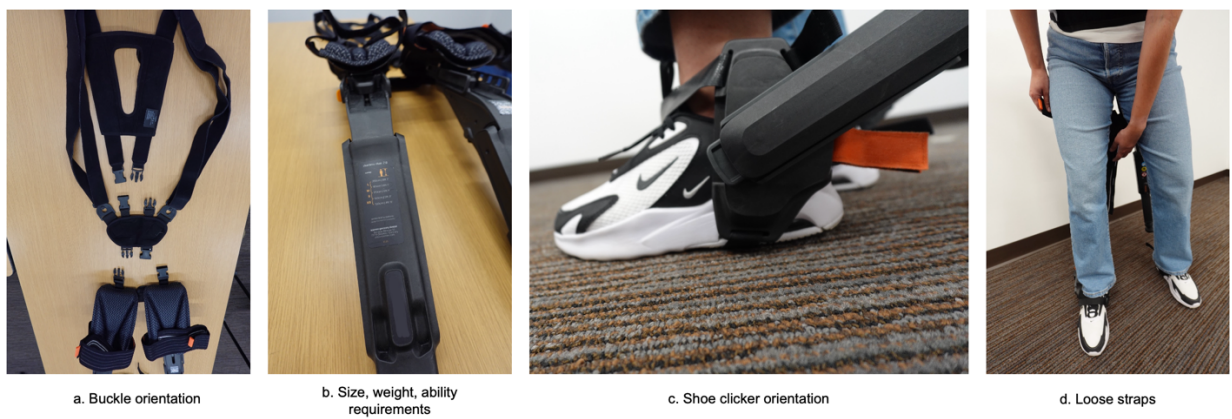

a. Buckle orientation    b. Size, weight, ability requirements    c. Shoe clicker orientation    d. Loose straps

**Figure 14a, b, c and d**. Major usability problems for the shoulder support exoskeleton organized from highest mean severity rating to lowest mean severity rating (left to right).

### *3.4. General, repeated major usability problems*

Table 7 shows the major usability problems, that are general to all exoskeletons, and repeated across exoskeletons by task phases. Assembly task had the largest number of general, repeated, major usability problems across the 3 exoskeletons, with 9 instances of major problems across all 3 exoskeletons. Evaluators rated poor documentation and a lack of sequence indicators as major problems for each of the 3 exoskeletons (Table 7) during assembly.

Accommodating different people was a general, repeated, major usability problem in all three exoskeletons across the four tasks – this was a problem in the assembly, donning and doffing tasks with the shoulder support exoskeleton, but was also a problem in the donning and disassembly tasks with the sit-stand exoskeleton, and the disassembly task in the back support exoskeleton. That the tasks required two people was another general, repeated major usability problem, with the shoulder support device requiring two people in the assembly and doffing tasks, the sit-stand and back support devices requiring two people in the donning task, and all three devices requiring two people in the doffing task.

**Table 7.** General (2.5 to 3.5 average severity ratings) repeated major problems across the 3 exoskeletons by task phases. [S] denotes the shoulder support exoskeleton; [SS] denotes the sit-stand device; [B] denotes the back support exoskeleton.

| General major problem | Assembly | Donning | Doffing | Disassembly |
|---|---|---|---|---|
| Accommodating different people | 1 [S] | 2 [S, SS] | 1 [S] | 2 [SS, B] |
| Two person operation | 1 [S] | 2 [SS, B] | 3 [S, SS, B] | |
| Documentation | 3 [S, SS, B] | | | |
| Sequence indicators | 3 [S, SS, B] | | | |
| Hazards/safety | | 1 [S] | 2 [S, SS] | |
| Manual strength requirements | 1 [SS] | 1 [S] | 1 [S] | 1 [S] |

Table 8 depicts the general, repeated, major usability problems and a count of heuristics these problems violated across exoskeletons and task phases. Universal design for usability, user control and freedom, and visibility of system status heuristics had the top three violations with 21, 18, and 15 violations, for the general, repeated, major usability problems identified across the three exoskeletons. Universal usability heuristic had the highest counts of violations, with 6 violations each for accommodating different people and for the task being a two person operation.

Requiring two people to be able to successfully operate this device was the general, repeated, major usability problem that violated the most heuristics, with a total of 25 violations of seven usability heuristics (out of the 14) across the three exoskeletons. Manual strength requirements also violated seven usability heuristics, with 14 total violations across the three exoskeletons.

**Table 8.** General major repeated problem and count of heuristics violated.

| Heuristic violated | Accommodating different people | Two person operation | Documentation | Sequence indicators | Hazards/safety | Manual strength required |
|---|---|---|---|---|---|---|
| Visibility | 3 | 4 | 2 | 2 | 3 | 1 |
| Match | 1 | 3 | 2 | 2 | | |
| User control | 2 | 5 | 1 | 3 | 3 | 4 |
| Consistency | 1 | | 1 | 2 | 2 | 1 |
| Error prevention | 2 | 3 | 1 | 1 | 3 | 2 |
| Recognition not recall | 1 | | | 3 | 3 | 1 |
| Flexibility | | 1 | | 1 | 1 | |
| Aesthetics | 1 | | 1 | 1 | | |
| Error recovery | 1 | | | | 1 | |
| Help & documentation | 1 | | 3 | | | |
| Universal usability | 6 | 6 | 2 | | 3 | 4 |
| Feedback | | | | 1 | | 1 |
| Closure | 1 | 3 | | 3 | 1 | |
| Language | | | 1 | 1 | | |

## 4. Discussion

Exoskeletons can become an important ally for industries to alleviate occupational injuries, but for industries to widely adopt them, and for these devices to become a routine part of daily work activities, their designs must be usable so that workers can feel comfortable wearing them for extended periods of time when performing their daily work tasks.

Furthermore, exoskeletons must be designed so that they are easy to assemble, don, doff, disassemble, and store in practice; otherwise, industries may balk at investing in these devices if these critical tasks cut into productive work time.

Therefore, to investigate whether the designs of the exoskeletons were usable, particularly when assembling, donning, doffing, and disassembling them, we conducted a heuristic design evaluation of the usability of three commercially available occupational exoskeletons. We used a combination of Nielsen and Shneiderman usability heuristics and evaluated a shoulder support, a back support and a sit-stand support exoskeleton to identify usability problems, particularly any catastrophic or major problems in the design of these devices when assembling, donning, doffing, and disassembling these devices.

Our key findings are as follows:

1. None of the three exoskeletons we evaluated had any catastrophic usability problems (severity rating of 3.5 to 4 on a 4-point scale), but all three exoskeletons had cosmetic (0 to 1.5 severity rating), minor (1.5 to 2.5 severity rating), and most importantly, major usability problems (2.5 to 3.5 severity rating). The shoulder support device had the most major usability problems compared with the back support and sit-stand support exoskeletons. Some of the unique major problems in the shoulder support exoskeleton include difficulties in adjusting belts, limited visual access, and high force requirements for activating the cup that held the arms securely, leading to significant safety hazards. The back support exoskeleton had an excessive number of parts, all of which had to be sorted and assembled with little help from the manual. In addition, sizing and fitting and tedious fastening and unfastening methods are unique problems with the device. In the sit-stand exoskeleton, the orientation of buckles, belts, and shoe clickers are major problems that can lead to fall hazards.

2. The most violated usability heuristics were visibility of system status, match between system and the real world, user control and freedom, error prevention, universal usability, and closure in task. Many displays and controls in the exoskeletons were out of visual access; hence, visibility to confirm if control actions worked, and, user control and freedom, error prevention, and achieving closure in tasks, were all impacted during the evaluation. The usability of these devices from the perspective of recognizing the needs of diverse users, such as novices to experts, different age ranges, disabilities, international variations in anthropometry and strength, and educational attainment, and accommodating their needs, was found to be problematic and could impact a wide variety of workers.
3. In all three exoskeletons evaluated, the maximum number of usability heuristic violations for major usability problems occurred during the assembly task phase.
4. Because different exoskeletons are designed to specifically support different core groups of muscles for functional effectiveness, we expected that any usability problems we found in any of the three exoskeletons would be unique and specific only to that exoskeleton. However, surprisingly, we found major usability problems that were general problems that repeated in more than one exoskeleton or in more than one task phase, potentially pointing to the pervasiveness of design weaknesses. The general, repeated and major usability problems included poor accommodation for different people pointing to sizing and fitting problems; assembly, donning, and doffing tasks requiring two people; poor documentation on how to assemble the devices; absence of good sequence indicators during assembly; safety hazards during donning and doffing; and high manual strength requirements, particularly for women and persons with disabilities.

## *4.1. Unique major usability problems in shoulder, sit-stand, and back support exoskeletons*

Adjusting the belt size in the harness during the assembly of the shoulder support exoskeleton is a major usability problem. There is a belt size indicator window along with line markings but a display of one size (number 45) is obscured in view by the Velcro straps and is not easy to locate visually to indicate any adjustments to the belt sizes that users can make. Furthermore, the belt size adjustment control itself is not easy to activate, and actions such as pulling, pushing, and tugging the adjustment control are not easy to discern because the control lacks any visual shape-based cues as to how users should activate it to adjust the size.

An additional major usability problem in the shoulder support device is the lack of unobstructed visual access to assembly points in the device. The slots in the harness, where the upper frame of the exoskeleton would fit and assemble, and hence, the most critical component in the exoskeleton for assembly of the entire device, are obstructed from direct visual access; they are located in the bottom back portion of the harness, and users cannot easily see these slots and locate them just by sight, especially if one were to fit them while donning. Furthermore, the slots are unremarkable in appearance, blend into the background, and do not contrast with the colors of the harness. In addition, there is no shape coding for visual recognition. Because visual access is obstructed, assembly cues that users can obtain visually by simply looking at the component parts and how they mate (that the frame rods would assemble into the harness slots in the back) are also difficult to perceive. The evaluators commented that when adjusting the harness around their waist, they accidentally touched the slots in the back and then used that knowledge to attempt to fit the frame tubes that connected the vest to the upper part of the exoskeleton. The question arises as to whether aesthetic considerations to make the exoskeleton appear homogenous, coherent, and uniform conflict with identifying important user-exoskeleton interaction points. In future work, identifying, highlighting, and analyzing user-exoskeleton interaction points that are critical for assembly, donning, doffing, and disassembly and developing iterative design recommendations might help address problems such as poor visual access to user-device control points. Our findings about a lack of visual access reinforce experiences from other researchers where a lack of visibility contributes to problems in reading the settings, which also wear out with use (S. A. Elprama et al., 2023).

The significant weight of the frames and arms create an additional problem in the shoulder support device, as users must hold these and maneuver them into the slots in the harness on their back side. Our work adds to the findings from other researchers who have recognized the problem of device weight and how it might impact wearing comfort for the user (Baltrusch et al., 2021; Cha et al., 2019; S. A. Elprama et al., 2023; Ferreira et al., 2020; Gilotta et al., 2019; Kim et al., 2019; Maurice et al., 2020; Satyajit Upasani & Srinivasan, 2019; Smets, 2019), but our finding is unique because the device weight can be a problem for the user not only for comfort after wearing it, but also in handling and manipulating the device even when assembling and donning it. Weight can become a difficult design choice involving material selection, tradeoffs between exoskeleton stiffness and strength, costs, and heaviness for the end user. However, exploring lighter weight materials and providing ways for users to hold and rest frames and arms while assembling and donning can resolve the problem.

The fastening and locking actions required of the user are also difficult to perform in the shoulder support device; actions include twisting, turning and rotating, pushing, pulling, pressing down, and then locking in many parts, including the frame assembly and arm cup assembly. These combined actions require excessive force application in one continuous movement of the user's hands and arms for actions to be effective. Breaking up the required user actions into discrete smaller steps instead of one large continuous step can provide the user with micro breaks for their hands and fingers during these actions. Furthermore, user hand and finger anthropometry, strength, and dexterity must be considered when designing the control and control surfaces so that users can obtain sufficient leverage and mechanical advantage when using these critical controls, and their actions can become less cumbersome and tedious. Although there is research documenting whole body fit of the exoskeleton to a user (Cha et al., 2019; S. Elprama et al., 2022; S. A. Elprama et al., 2023; Gilotta et al., 2019; Kim et al., 2019; Smets, 2019), there is no research documenting specific user-control interaction points and the anthropometric fit of extremities such as the fingers and the hand needed for precise control activation.

      The sit-stand exoskeleton is typically worn either around the waist in a waist harness or in tandem with a shoulder harness, with the waist and shoulder harnesses connected through a system of belts, mating buckles, and Velcro with a multitude of adjustments that can be made to fit the user. Orienting the buckles in the device so that proper connections between the waist and shoulder harnesses can be achieved is a major usability problem. First, it is unclear whether a shoulder harness is required. Second, the direction and orientation of the mating buckles in the shoulder and waist harnesses are not distinct and clear, so how the shoulder harness would interface with the waist harness and connect to the seat with the belts, buckles, and Velcro is confusing. Given the large number of belts, buckles, and Velcro in the device, visual (even color-coded) orientation and direction markers could solve this problem. The device also consists of a shoe clicker that connects to a leg frame with an attached seat – the shoe clicker, which is worn around the shoe with a strap, is difficult to orient to a person's shoe because it is not shaped like a shoe. Furthermore, the clicker must be worn first before the leg frames – after wearing the leg frames, users will not be able to bend or sit and wear the clicker. While many elements of the device conform to the human form such as the lower leg frame of the exoskeleton that resembles a person's legs, and the seat that has a form that

indicates that it is clearly meant for sitting, important connecting components that hold the exoskeleton together, such as the shoe clicker, are odd-shaped and do not visually and intuitively represent a form or follow the form of the feet so that the user can relate to the shape immediately. The necessity and relevance of this odd shape for shoe clickers is unclear. The shape is even more important, given that the clicker must be worn first for the sequence of donning to work effectively without safety hazards.

    Sizing adjustments are placed on the side of the sit-stand exoskeleton, and once donned, if adjustments must be made, the device must be completely doffed to make those them, or a second person will be needed to help with the adjustments – the controls are located too distant from the user so adjusting sizes is tedious and effortful. One cannot make any size adjustments in the seated position. In doffing the device, the device extends all the way to the shoes and feet; taking the shoe clickers and the straps off requires tedious bending and balancing actions and can result in a fall hazard. These design problems violate the principles of human factors that recommend placing controls within the user's reach. Our findings add to other work (S. A. Elprama et al., 2023) that concludes that users must be able to reach belts and clips for any needed adjustments by themselves without having to rely on help from others.

    The back support exoskeleton is a complex design with many different unique parts making it overwhelming to assemble, and difficult to intuitively see how the component parts fit together.  Many of the parts are meant for either different sizes or as optional pieces. To reduce the burden of having the user comprehend these parts, become overwhelmed in the process, and feel a lack of control, manufacturers can package the parts better in their own shape-fitting boxes and label them clearly so that users can immediately recognize the parts they need to assemble according to their sizes. Future work should also consider how portability of the device, where every part can be disassembled and put in a bag to carry, and

the number of choices and options in the device, trade off in design with the complexities of having too many individual parts.

Other researchers have related the parts packaging problem to concerns such as storing the loose parts so they are not lost (S. A. Elprama et al., 2023), and for ensuring compactness of the device for storage (Cha et al., 2019; Omoniyi et al., 2020; Smets, 2019), but our concern with packaging is based on a central problem of users feeling a lack of control and being overwhelmed with all the loose parts in a bag. Sizing and fit are problematic in the back support exoskeleton; fitting while donning the device requires assistance from a second person especially because neither does the design indicate and provide feedback on whether one is wearing the device incorrectly, nor does it indicate how one is supposed to "feel" after donning it. Our findings about fit add further evidence to the numerous studies that have documented the need for a good fit considering gender, body shapes and sizes (Cha et al., 2019; S. Elprama et al., 2022; S. A. Elprama et al., 2023; Gilotta et al., 2019; Kim et al., 2019; Smets, 2019). Fit is a critical component of the exoskeleton if it is to provide the required support during an industrial task, and because fitting the back support exoskeleton requires an additional person, implementing this in industry can pose a significant barrier in both compliance and time required.

*4.2. Heuristics violated the most*

In the three exoskeletons evaluated, the visibility of the system status, user control and freedom, error prevention, universal usability, task closure, and the match between the system and the real world, were the most violated heuristics. Users must primarily rely on information present in the visual labels, signs, and cues on the exoskeleton to take control actions to either assemble the devices or to don them which becomes problematic, particularly if the user wants to know if their actions work in securing the parts during donning. For instance, the back support exoskeleton contains key information, such as sizing only in visual form and in the form of a color-coded sizing guide in the manual. Two of the exoskeletons evaluated, shoulder support and back support devices, require the user to wear the devices on the back. Users can either completely assemble these devices and then don them or they can choose to assemble a few parts in the device while donning the device; the designs themselves do not have any stop constraints in preventing users from choosing one of these two options. However, regardless of what users choose to do, the actions needed to activate controls in the devices still require the user to exert force in the absence of visual access – it is just that these actions occur separately during assembly or jointly when assembly overlaps with donning. For instance, securing the flex frames and the arm cups in the shoulder exoskeleton requires not only significant force exertion to ensure that the mechanisms lock together, but also that this force must be exerted in actions such as twisting, pushing, pinching with thumbs, and rotating the hand in the absence of unobstructed visual access to the control. This presents challenges not only in visually checking for fit but also in preventing the user from ensuring closure in the task. Hence, due to the lack of visibility, adjustments to the exoskeletons must be made either with a second person helping or through trial and error, which often results in users having to doff, disassemble, and retrace their steps back.

    All three exoskeletons could do better in providing the user with a sense of control during assembly, donning, and doffing tasks. For example, the back support exoskeleton has too many unique parts, and when beginning to assemble the device, users would feel the tediousness involved and become overwhelmed, and would be unable to obtain a sense of initial orientation for the parts and for how, once assembled, the device would be configured to support the back. Activating controls such as torso structures in conjunction with the smart joints in the hip frame and how the knee pads would support the movement of the back is not

intuitive during assembly, resulting in a lack of a sense of control over how to assemble this complex device. The critical user-device interaction points, especially the controls users interact with in ensuring the fit and safety of the device, such as interlocking mechanisms between two safety-critical parts, need careful rethinking in design.

In the shoulder support exoskeleton, the lack of user control and freedom seems to come from the myriad of assembling and donning actions that the user must activate with the control elements and surfaces in the exoskeleton. For instance, the flex frame slots are in the back of the harness, making it difficult to locate and easily discover making assembly challenging. The force requirements and locking actions with the arm cup where the upper arms would rest when using the device and provide shoulder support are effortful and tedious, and take many attempts before the user can succeed in locking the parts together so that the parts will not come off. The design intent may have been to prevent accidental activation and unlocking, and as a fail-safe mechanism, designers may have intended for the user to push, twist, and turn for the device to lock, but this makes users exert more force than what may be needed with the multitude of actions they must take and be effective at, for the mechanism to work. Exoskeletons are intended to reduce pain and exertion when performing an industrial task (S. A. Elprama et al., 2023; Satyajit Upasani & Srinivasan, 2019), so they should not inadvertently be the cause for pain and discomfort due to the force requirements and tedious effort involved in assembling and donning the device.

In the sit-stand exoskeleton, a lack of user control is particularly felt when attaching leg frames to shoe sliders and clickers. This lack of control is especially accentuated during donning when the user tries to sit on the exoskeleton seat and balance using lower leg frames. Added vigilance is required by the user because if the fit is not adjusted appropriately, there is potential for the user to lose balance and fall with the device still attached to them.

A usability violation is a lack of tolerance for user errors during assembly and donning, and the user having to retrace their steps back and redo their tasks. A lack of visibility of the system status, added to the lack of user control, and a lack of a sense of closure on whether the task was completed contributes to the potential for user errors. However, the devices themselves carry little or no warning signs for user errors – the written user manuals do offer instructions on how to assemble, don, and doff, but when users make an error in a step, there is no warning from the device about the error, and the user will not realize that they made an error until after a few more steps when the device either does not fit together or does not function as expected. All three exoskeletons we evaluated were not active exoskeletons, so how error warnings can be designed for a passive exoskeleton is a design challenge. Furthermore, constraints for users based on both sequence and part fitting and mating should be incorporated into the designs so that users cannot take the next steps if they commit an error. Extensive user testing and error-proofing can also be conducted during the early design stages to root out the use errors as much as possible.

Our findings also suggest that universal usability is a major problem in the exoskeletons as these devices do not adequately consider all users and use conditions in their design (Martinez et al., 2024). Many actions required for assembly and donning in exoskeletons can be difficult for users with diverse abilities. For instance, many assembly and donning actions in exoskeletons rely on the use of sight (visual mode), with limited reliance on auditory (clicks when assembling) and tactile modes (locating slots by touch and shape). Given the large number of similarly shaped parts for assembly and that many of the instructions in the manual are visual, workers with sensory limitations will find it difficult to perform many of the critical assembly and donning tasks. Safety concerns such as with the sit-stand exoskeleton or shoulder support exoskeleton's arm bouncing back can be exacerbated in people with balance problems or those with limited mobility. These devices are also not usable by workers with diverse abilities, users with canes, or other assistive devices. Universal design principles must be considered when designing exoskeletons so that all workers, regardless of their age, gender (S. A. Elprama et al., 2023), ability, functional status, and other characteristics, can access, use, and benefit from them.

Finally, because the concept of the exoskeletons is still new, and because the designs, the parts, the terminology used, and the general forms of the devices are new, complex and unique, designers of these devices could strive for a better match with the real world, giving users a better intuitive understanding of how specifically the design of the devices would achieve the musculoskeletal support function, and what benefits would result through user training materials and through the design of the devices. Studies have shown that if workers perceive exoskeletons as useful and helpful based on the physical support they receive from the device, and if they gain more knowledge of exoskeletons and how they work, they are more likely to adopt these devices for their work tasks (S. A. Elprama et al., 2023; Ferreira et al., 2020; Gilotta et al., 2019).

### *4.3. Task phases where most usability problems occurred*

The assembly and donning tasks present the most major usability problems, with average severity ratings between 2.5 and 3.5. Users require time and effort to develop a visual map of the individual parts and how they fit and to refer to the manual to understand the sequence of steps required for assembly. Assembly tasks also involve actions from the user that demand force, strength, mobility, and balance. If exoskeletons cannot be assembled, they cannot be worn. Furthermore, if the industry is to adopt exoskeleton technology, the task of assembling the exoskeletons must be made simpler and quicker, and making the design more usable during assembly can help achieve these goals. Therefore, our findings highlight the importance of designing the exoskeleton for ease of use when assembling it. Even if one were to pre-assemble some parts of the exoskeleton and have it ready, some final assembly steps may still be required before and during the donning task, so it is critical for the exoskeleton to have a high degree of usability with minimal problems during assembly. Furthermore, how a device is assembled can also affect how the device is doffed when workers must take breaks or doff the exoskeleton at the end of the workday.

Additionally, some exoskeletons, such as the sit-stand device, contain overlaps between the assembly and donning tasks for the user to complete an effective sequence. These overlap points pose safety risks; however, the design does not include any stop constraints to protect users from these risks.

Our finding that many usability violations occur during assembly or donning is concerning given that one must progress through these stages to get the exoskeleton ready for use. This finding highlights the urgent need for designers and manufacturers to consider and design these pre- and post-use tasks so that users can succeed in easily and quickly completing them. Future research should examine how the factors related to industry adoption and worker compliance are affected by assembly, donning, doffing, and disassembly.

Although the doffing and disassembly stages do not pose as much concern as the assembly and donning stages do, they pose safety challenges. Some doffing actions require users to be vigilant to prevent safety hazards. For example, removing the sit-stand exoskeleton requires bending to detach the feet that are still connected to the entire exoskeleton on the person. Similarly, detaching arm cuffs from the shoulder support exoskeleton results in a significant bounce-back and ricochet of a heavy metal arm, introducing a significant safety hazard when the user could get hit in the face with the metal arm of the exoskeleton. Some disassembly tasks, also required strength to detach parts in some exoskeletons; sometimes they required using both hands to press down on levers to release the parts, not only requiring force, but also introducing pinch hazards. In most cases, doffing and disassembly required steps in the reverse order of assembly and donning, which explains the fewer violations of criteria in these two stages compared to assembly and donning.

### *4.4. General, repeated major usability problems*

Although the three exoskeletons evaluated in the study were all different in their designs to support a different muscle group, surprisingly, we found major usability problems repeated across the three exoskeletons we evaluated. These major problems included weaknesses in accommodating different people, all three exoskeletons requiring two people to operate, poor documentation in the instruction manuals, the absence of sequence indicators in these devices, safety hazards when both donning and doffing these, and manual strength requirements during all task phases.

Accommodating different user groups, particularly, women workers, older workers, workers who may have reduced mobility, strength, and ranges of motion, and workers with other forms of disabilities would be problematic with all these exoskeletons. These concerns also exist in all four task phases. For instance, assembling the shoulder support device requires not only gross manual strength but also fine fingertip dexterity and strength for assembling certain controls such as the flex frame-harness assembly and the arm cup lock, both of which require pushing, twisting, and rotational actions from the user. With the back support exoskeleton, because the device is heavy, doffing, particularly the hip frame and belt assembly, would be tedious and difficult for women, older workers, and workers with disabilities. The major problem is that the designs require users to use significant hand and finger strengths for control activation. Actions such as unscrewing components apart can easily be redesigned to include other types of controls, such as push buttons to disengage components. The sit-stand device is heavy to handle, and because there is considerable overlap between assembly and donning tasks, donning the leg frames in the device requires additional assistance for people with reduced mobility and strength. Users with disabilities would find it difficult to progress beyond assembly task. When designing, testing, and evaluating industrial exoskeletons, manufacturers must recruit and consider diverse users so that their testing samples are representative of the industrial workforce, which consists of a wide variety of workers.

Another major usability problem that was repeated across the three exoskeletons was that these devices required one additional person (two-person operation) for successful assembly, donning, and doffing. This problem of needing two persons for donning, doffing and adjusting has been recognized by Elprama et al. (2023) based on their own experience. The weight, volume, and awkward shapes of the devices when partially assembled impact the handling of the devices by just one person. For instance, for donning the shoulder support device, iterative checking of fit is required, but the fit check requires two persons. The donning of the device also requires that the wearer assume postures that can become awkward when wearing it like a backpack, but also when fitting and adjustments need to be made, the force and strength requirements necessitate assistance from a second person. Having an additional person help with donning, fitting and doffing these devices can pose a significant barrier in industry, given the impacts additional labor and time would have on production requirements in industry.

All three devices had poor documentation, particularly with respect to instructions on the sequence of assembly steps. Documentation can significantly impact how workers are trained to use exoskeletons, and how they learn to use it (S. A. Elprama et al., 2023; Gilotta et al., 2019; Moyon et al., 2019; Spada et al., 2017). In the documentation for the sit stand exoskeleton, many of the instructions are in the form of pictures, but the pictures are in the form of line drawings. Including only pictures is not useful without supplementary information in the form of instructional steps to support the pictures. Documentation has room for improvement; creating dedicated multimedia resources and creating a community forum for users to share their experiences can help enhance learning how to use these devices.

All three exoskeletons could include sequence indicators for the assembly task. It is difficult to figure out sequences simply by looking at the overall assembled pictures in the manual. It is also important for the designs to include a clear beginning, middle, and an end in the sequence of assembly tasks, because it is easy to miss connections between the phases, which is especially pronounced for the shoulder support device and the sit-stand device, both of which have significant problems with visual access, especially when the user must either look at the back of the person for the assembly points or must bend and look down for the assembly points. In addition to providing clear sequence indicators on the device, designing in sequence and part constraints so there is only one way to fit and assemble the parts can address these challenges.

A major usability problem that repeated with shoulder and sit-stand devices is the safety hazards posed by these two devices. Elprama et al. (2023) and Näf et al. (2018) have recommended that moveable exoskeleton parts should not touch or bump into each other, but we encountered a different, more critical type of a collision risk. The heavy metal arm in the shoulder support exoskeleton bounces back and ricochets, introducing a significant safety hazard, whereby the user can get hit in the face. The sit-stand exoskeleton can cause the user to lose balance and fall back, particularly when donning and doffing the device. Other researchers (Kim & Nussbaum, 2019; Satyajit Upasani & Srinivasan, 2019) have also acknowledged the potential for fall risk with exoskeletons, so our findings add evidence to this potential risk. Additionally, both devices pose significant hazards from numerous pinch points, sometimes with sharp edges that users must carefully avoid. It is not clear if a thorough safety and hazard evaluation is conducted during the design of these exoskeletons and if the results from the evaluation are fed back to improve the design and eliminate hazards. Future research studies in exoskeletons should include safety and hazard evaluations in addition to performance assessments of these devices.

## 5. Study limitations

We evaluated only three exoskeletons in our study, in a growing market of occupational exoskeletons. Even though our findings about extremely specific design features in the exoskeletons we evaluated may not extend exactly to other devices in the market, the devices we evaluated, especially the back support and the shoulder support devices, have been extensively researched by others. Even though we only evaluated three specific exoskeletons, we surprisingly found general design weaknesses that led to major usability problems that repeated across the three exoskeletons – this indicates that some usability and design problems may be generalizable across any exoskeleton, and needs further investigation. We also chose to evaluate one back support device, one shoulder support device, and one sit-stand device. The devices, therefore, cover important body segments that could be subject to musculoskeletal exertions – the back and lumbar flexion, and the shoulder region for upper extremities being body segments of common concern and traditional focus for ergonomists attempting to study and solve the problem of musculoskeletal injuries in the workplace. The sit-stand exoskeleton could also have wide ranging applications in many occupations, and our usability evaluation could prove useful for any redesigns of this device.

Although our study did not involve end users (workers), and although we did not study exoskeleton use during an industrial task, our evaluation is a design evaluation of the exoskeletons providing information about granular, specific design weaknesses. Design evaluations focus on specific design features and is comprehensive using usability heuristics. Our goal was also to see if there are design weaknesses across exoskeletons at the user-device interaction points and how that interaction is typically achieved. Our focus was also on four tasks that are important before and after actual use of the devices – that is, pre-use and post-use tasks including assembly, donning, doffing, and disassembly of the exoskeletons.

## 6. Conclusions

For large scale industrial implementation of exoskeletons to succeed so the benefits of this innovation can be reaped by industry, our view is that critical product design elements and features such as whether the device is portable and storable, is simple and parsimonious in design with minimal number of parts, is easily cleaned and sanitized, is maintainable and repairable, is safe without any hazards where a worker could get injured trying to don or doff it, in addition to its quality and reliability over the long run must be investigated further. From the perspective of the industry, we think many critical factors will impact adoption of exoskeletons including the following: (1) space requirements and constraints for storing these devices - some devices like the sit-stand exoskeleton are small and fit in a small case; others that support the shoulder and the back are larger and need more storage space; (2) how quickly they can be put on and taken off with minimal to no training burden much like regular personal protective equipment such as hard hats and safety glasses; and, (3) most importantly, the costs and benefits for industries to widely implement these for every worker.

We recommend the following future directions for research: (1) evaluations of exoskeletons must include in-depth safety, accessibility and universal design evaluations, given we found significant safety hazards and poor accommodation for a diverse user group, especially potential older workers, workers with potential mobility impairments, and women; (2) future research should investigate the design and implementation factors that can qualify exoskeletons as a personal protective equipment.

Daratany, C., & Taveira, A. (2020). Quasi-experimental study of exertion, recovery, and worker perceptions related to passive upper-body exoskeleton use during overhead, low force work. *Human Interaction, Emerging Technologies and Future Applications II: Proceedings of the 2nd International Conference on Human Interaction and Emerging Technologies: Future Applications (IHIET–AI 2020), April 23-25, 2020, Lausanne, Switzerland*, 369–373.

De Vries, A. W., Krause, F., & De Looze, M. P. (2021). The effectivity of a passive arm support exoskeleton in reducing muscle activation and perceived exertion during plastering activities. *Ergonomics*, *64*(6), 712–721. https://doi.org/10.1080/00140139.2020.1868581

Elprama, S. A., Bock, S. D., Meeusen, R., Pauw, K. D., Vanderborght, B., & Jacobs, A. (2023). Design and Implementation Requirements for Increased Acceptance of Occupational Exoskeletons in an Industrial Context: A Qualitative Study. *International Journal of Human–Computer Interaction*, 1–16. https://doi.org/10.1080/10447318.2023.2247597

Elprama, S., Vanderborght, B., & Jacobs, A. (2022). An Industrial Exoskeleton User Acceptance Framework Based on a Literature Review of Empirical Studies. *Applied Ergonomics*. https://doi.org/10.1016/j.apergo.2021.103615

Elprama, S., Vannieuwenhuyze, J., Bock, S. D., Vanderborght, B., Pauw, K. D., Meeusen, R., & Jacobs, A. (2020). Social Processes: What Determines Industrial Workers' Intention to Use Exoskeletons? *Human Factors the Journal of the Human Factors and Ergonomics Society*. https://doi.org/10.1177/0018720819889534

Ferreira, G., Gaspar, J., Fujão, C., & Nunes, I. L. (2020). Piloting the use of an upper limb passive exoskeleton in automotive industry: Assessing user acceptance and intention of use. *Advances in Human Factors and Systems Interaction: Proceedings of the AHFE 2020 Virtual Conference on Human Factors and Systems Interaction, July 16-20, 2020, USA*, 342–349.
49

Kim, S., & Nussbaum, M. A. (2019). A Follow-Up Study of the Effects of An Arm Support Exoskeleton on Physical Demands and Task Performance During Simulated Overhead Work. *IISE Transactions on Occupational Ergonomics and Human Factors*, 7(3–4), 163–174. https://doi.org/10.1080/24725838.2018.1551255

Kim, S., Nussbaum, M. A., Esfahani, M. I. M., Alemi, M. M., Alabdulkarim, S., & Rashedi, E. (2018). Assessing the influence of a passive, upper extremity exoskeletal vest for tasks requiring arm elevation: Part I - "Expected" effects on discomfort, shoulder muscle activity, and work task performance. *Applied Ergonomics*. https://doi.org/10.1016/J.APERGO.2018.02.025

Klarich, A., Noonan, T. Z., Reichlen, C., Barbara, S. M. J., Cullen, L., & Pennathur, P. R. (2022). Usability of smart infusion pumps: A heuristic evaluation. *Applied Ergonomics*, 98, 103584. https://doi.org/10.1016/j.apergo.2021.103584

Kuber, P. M., Abdollahi, M., Alemi, M. M., & Rashedi, E. (2022). A Systematic Review on Evaluation Strategies for Field Assessment of Upper-Body Industrial Exoskeletons: Current Practices and Future Trends. *Annals of Biomedical Engineering*. https://doi.org/10.1007/s10439-022-03003-1

Kuber, P. M., Alemi, M. M., & Rashedi, E. (2023). *A Systematic Review on Lower-Limb Industrial Exoskeletons: Evaluation Methods, Evidence, and Future Directions*. https://doi.org/10.1007/S10439-023-03242-W

Liu, S., Hemming, D., Luo, R. B., Reynolds, J., Delong, J. C., Sandler, B. J., Jacobsen, G. R., & Horgan, S. (2018). Solving the surgeon ergonomic crisis with surgical exosuit. *Surgical Endoscopy*, 32(1), 236–244. https://doi.org/10.1007/s00464-017-5667-x

Mack, R. L., & Nielsen, J. (1995). Usability inspection methods: Executive summary. In *Readings in human–computer interaction* (pp. 170–181). Elsevier.
51